# Turbulent Heating in Galaxy Clusters Brightest in X-rays

I. Zhuravleva[1,2], E. Churazov[3,4], A. A. Schekochihin[5,6], S. W. Allen[1,2,7], P. Arévalo[8,9], A. C. Fabian[10], W. R. Forman[11], J. S. Sanders[12], A. Simionescu[13], R. Sunyaev[3,4], A. Vikhlinin[11] & N. Werner[1,2]

[1]KIPAC, Stanford University, 452 Lomita Mall, Stanford, CA 94305, USA

[2]Department of Physics, Stanford University, 382 Via Pueblo Mall, Stanford, CA 94305-4060, USA

[3]Max Planck Institute for Astrophysics, Karl-Schwarzschild-Strasse 1, D-85741 Garching, Germany

[4]Space Research Institute (IKI), Profsoyuznaya 84/32, Moscow 117997, Russia

[5]The Rudolf Peierls Centre for Theoretical Physics, University of Oxford, 1 Keble Rd, Oxford OX1 3NP, UK

[6]Merton College, Oxford OX1 4JD, UK

[7]SLAC National Accelerator Laboratory, 2575 Sand Hill Road, Menlo Park, CA 94025, USA

[8]Instituto de Física y Astronomía, Facultad de Ciencias, Universidad de Valparaíso, Gran Bretana N 1111, Playa Ancha, Valparaíso, Chile

[9]Instituto de Astrofísica, Facultad de Física, Pontificia Universidad Católica de Chile, 306, Santiago 22, Chile

[10]Institute of Astronomy, University of Cambridge, Madingley Road, Cambridge CB3 0HA, UK

[11]Harvard-Smithsonian Center for Astrophysics, 60 Garden Street, Cambridge, MA 02138, USA

[12]Max-Planck-Institut für extraterrestrische Physik, Giessenbachstrasse 1, 85748 Garching, Germany

[13]Japan Aerospace Exploration Agency, 3-1-1 Yoshinodai, Sagamihara, Kanagawa 229-8510, Japan

**The hot ($10^7$ -$10^8$ K), X-ray-emitting intracluster medium (ICM) is the dominant baryonic constituent of clusters of galaxies. In the cores of many clusters, radiative energy losses from the ICM occur on timescales significantly shorter than the age of the system[1,2,3]. Unchecked, this cooling would lead to massive accumulations of cold gas and vigorous star formation[4], in contradiction to observations[5]. Various sources of energy capable of compensating these cooling losses have been proposed, the most promising being heating by the supermassive black holes in the central galaxies through inflation of bubbles of relativistic plasma[6-9]. Regardless of the original source of energy, the question of how this energy is transferred to the ICM has remained open. Here we present a plausible solution to this question based on deep Chandra X-ray observatory data and a new data-analysis method that enables us to evaluate directly the ICM heating rate due to the dissipation of turbulence. We find that turbulent heating is sufficient to offset radiative cooling and indeed appears to balance it locally at each radius – it might therefore be the key element in resolving the gas cooling problem in cluster cores and, more universally, in atmospheres of X-ray gas-rich systems.**

Perseus and Virgo/M87 are well studied nearby cool-core clusters of galaxies in which the central cooling times, due to the emission of X-rays, are an order of magnitude shorter than the Hubble time (Methods, Extended Data Fig. 1). X-ray observations show that the ICM in central regions of these clusters is disturbed, suggesting that it might be turbulent. The most likely drivers of this turbulence are mechanically powerful active galactic nuclei (AGN) in the central galaxies of both clusters, which inflate bubbles of relativistic plasma in the ICM. During the inflation and subsequent buoyant rise of these bubbles, internal waves and turbulent motions in the gas can be excited[10,11,12], which must eventually dissipate into heat. In order to determine whether this heating is sufficient to balance radiative losses and prevent net cooling, one must estimate the turbulent heating rate – and for that, a measurement is needed of the rms (root mean square) turbulent velocity amplitude $V$ as a function of length scale $l$. Then the turbulent heating rate in the gas with mass density $\rho$ is (dimensionally) $Q_{\text{turb}} \sim \rho V^3/l$, to within some constant of order unity that depends on the exact properties of the turbulent cascade. $Q_{\text{turb}}$ has never previously been probed directly mainly because of two difficulties. In this Letter we propose ways of overcoming both, leading to an observational estimate of $Q_{\text{turb}}$ and a tentative conclusion that it is sufficient to reheat the ICM.

The energy resolution of current X-ray observatories is insufficient to measure gas velocities in the ICM, or their dependence on scale. Here, we circumvent this problem by instead measuring gas density fluctuations and inferring from their power spectrum the power spectrum of the velocities. A simple theoretical argument, supported by numerical simulations, shows that in relaxed galaxy clusters, where the gas motions are subsonic, the rms amplitudes of the density and one-component velocity fluctuations are proportional to each other at each scale $l=k^{-1}$ within the inertial range[13,14]: $\delta\rho_k/\rho_0 \approx \eta_1 V_{1,k}/c_s$, where $\rho_0$ is the mean gas density, $c_s$ the sound speed and $\eta_1$ is the proportionality coefficient ~1 set by gravity-wave physics at large, buoyancy-dominated scales[13]. Here we define $V_{1,k}$ by $3V_{1,k}^2/2 = k_1 E(k_1)$, where $k_1=2\pi k$ is the traditional Fourier wave number and $E(k_1)$ is the energy spectrum of the three-dimensional velocity field; $\delta\rho_k/\rho_0$ is defined analogously in terms of the density fluctuation spectrum, but without the factor of 3/2. Un-sharp-masked images of the Perseus Cluster show ripple-like structures in the core, reminiscent either of sound waves[15,16] or stratified turbulence[13,17] (Methods). Here we investigate the consequences of the second scenario (which may be argued to be more likely if the stirring of the ICM by the AGN ejecta is of sufficiently low frequency).

The high statistical precision obtained by Chandra with a 1.4 Ms observation of the Perseus Cluster core makes this data set ideal for probing density structures over a range of spatial scales. Fig. 1 shows the mosaic image and a residual image, made by dividing the mosaic image by a spherically symmetric $\beta$ model of the mean intensity profile with core radius 1.26' ≈ 26 kpc and slope $\beta$=0.53 (Methods, Extended Data Fig. 2). Using the modified $\Delta$-variance method[18], we calculate the power spectra of surface-brightness fluctuations in a set of concentric annuli (Extended Data Fig. 3), each with width 1.5' (31 kpc), and deduce from them the amplitudes of density fluctuations across a range of spatial scales. The typical $\delta\rho_k/\rho_0$ at $k^{-1} \sim 20$ kpc varies from ~20% inside the central 1.5' (31 kpc) to ~7% at the distance of ~10.5' (218 kpc) from the cluster center (I.Z. *et al.*, manuscript in preparation). We have also performed a similar analysis for a ~600 ks Chandra observation of the M87/Virgo cluster.

Fig. 2 shows examples of the velocity amplitudes $V_{1,k}$ inferred from the density amplitudes $\delta\rho_k/\rho_0$ via the relation $\eta_1 V_{1,k}/c_s \approx \delta\rho_k/\rho_0$, in two different annuli for each of the two clusters. In these examples, over the range of spatial scales where the measurements are robust, $V_{1,k}$ varies from ~70 km s$^{-1}$ to ~145 km s$^{-1}$ in Perseus. In the full set of 7 annuli from the center to 10.5' (218 kpc), the range of velocities is larger, up to 210 km s$^{-1}$. In Virgo, the typical velocity amplitudes in all annuli are smaller, between 43 and 140 km s$^{-1}$, but the corresponding spatial scales are smaller too.

These (inferred) velocity spectra can be used to estimate the heating rate $Q_{\text{turb}} \sim \rho V^3/l$. The second difficulty mentioned earlier is that normally $l$ here is taken to be the energy-containing scale of the turbulence, which is difficult to determine or even define unambiguously: in theory, several characteristic scales (e.g., the distance from the center, various scale heights, etc.) are present in the problem[19]. The measured spectra (Fig. 2) do not necessarily offer clarity about the injection scale, since at low $k$ they are dominated by large-scale inhomogeneities and the radial width of the chosen annuli. However, in a turbulent cascade, the energy spectrum *in the inertial range* should have a universal form depending only on $k$ and the mean, density-normalized dissipation rate $\varepsilon = Q_{\text{turb}}/\rho_0$. Assuming purely hydrodynamic[20] turbulence, the energy spectrum should be $E(k_1) = C_K \varepsilon^{2/3} k_1^{-5/3}$, where the Kolmogorov constant[21,22] $C_K \approx 1.65$. The turbulent energy flux at any scale in the inertial range will be the same and equal to the mean dissipation rate: accounting for our convention $k=1/l=k_1/2\pi$ and $V_{1,k}=[2k_1E(k_1)/3]^{1/2}$, we obtain $Q_{\text{turb}}=\rho_0\varepsilon=C_Q\,\rho_0 V_{1,k}^3 k$, where $C_Q=3^{3/2}\,2\pi/(2C_K)^{3/2}\approx 5$ is a dimensionless constant whose value should be treated as a fiducial number. Indeed, while the constant-flux, Kolmogorov-like nature of the turbulence is probably a good assumption, the specific constant $C_Q$ will depend on more detailed properties of the turbulent cascade (e.g., magnetohydrodynamic rather than hydrodynamic[23]) and, in particular, on the types of fluctuations that carry the total injected energy flux to small scales (velocity, magnetic, density fluctuations[24]). We will not be concerned here with a precise determination of $C_Q$. It is clearly an order-unity number and it is also clear that our estimate for the turbulent heating rate can only be used if we identify, for each of the annuli where we calculated $V_{1,k}$, a $k$ interval in which $V_{1,k}^3 k$ stays approximately constant with $k$. Remarkably, our measured velocities are indeed consistent with $V_{1,k} \sim k^{-1/3}$, accounting for the errors and uncertainties associated with finite resolution and with our method of extracting power spectra[25].

Because of order-unity uncertainties in the determination of $Q_{\text{turb}}$, the question of heating-cooling balance boils down to whether the local $Q_{\text{turb}}$ measured at each radius is comparable within an order of magnitude to the local cooling rate and, more importantly, scales linearly with it from radius to radius and between clusters. The answer, as demonstrated by Fig. 3, is yes. Here the gas cooling rate was evaluated directly from the measured gas density and temperature $T$, $Q_{\text{cool}}=n_e n_i \Lambda_n(T)$, where $n_e$ and $n_i$ are the number densities of electrons and ions, respectively, and $\Lambda_n(T)$ is the normalized gas cooling function[26]. We see that, in all 7 annuli in Perseus and all 4 in Virgo (which span the cluster cores in both cases), $Q_{\text{turb}} \sim Q_{\text{cool}}$ over nearly three orders of magnitude in the values of either rate (Fig. 3, Methods). Note that in Virgo and Perseus similar levels of $Q_{\text{cool}}$ and $Q_{\text{turb}}$ are attained at physically different distances from the cluster centers.

While these results are encouraging, the uncertainties associated with the above analysis are, admittedly, large (Methods). It is difficult to prove unambiguously that we are dealing with a universal turbulent cascade, as other structures (e.g., edges of the bubbles, entrainment of hot bubble matter[12], sound waves[15,16], mergers and gas sloshing[27]) might also contribute to the observed density-fluctuation spectra. Rather we argue simply that the cluster cores appear disturbed enough that if these disturbances are indeed due to turbulence, then its dissipation can reheat the gas. At the very least, one may treat the amplitudes calculated here (Fig. 2) as an upper limit on the turbulent velocities. One of the major tasks for future X-ray observatories, capable of measuring the line-of-sight gas velocities directly, will be to verify the accuracy of these velocity amplitudes.

Modulo this caveat, the approximate balance of cooling and heating (Fig. 3) suggests that turbulent dissipation may be the key mechanism responsible for compensating gas cooling losses and keeping cluster cores in an approximate steady state. While AGN activity is not the only driver of gas motions (mergers or galaxy wakes can contribute as well[28]), it is plausible that AGNs play the dominant role in the central ~100 kpc, where the cooling time is short. If this is true, then our results support the self-regulated AGN feedback model[10], in which unchecked cooling causes accelerated accretion onto the central black hole, which responds by increasing the mechanical output, presumably in the form of bubbles of relativistic plasma; the bubbles then rise buoyantly, exciting in particular internal waves[11,29]; the energy from them is converted into turbulence, which cascades to small scales and eventually dissipates, reheating the gas.

1. Lea, S. M. The dynamics of the intergalactic medium in the vicinity of clusters of galaxies. *Astrophys. J.* **203**, 569-580 (1976).
2. Cowie, L. L. & Binney, J. Radiative regulation of gas flow within clusters of galaxies - a model for cluster X-ray sources. *Astrophys. J.* **215**, 723-732 (1977).
3. Fabian, A. C. & Nulsen, P. E. J. Subsonic accretion of cooling gas in clusters of galaxies. *Mon. Not. R. Astron. Soc.* **180**, 479-484 (1977).
4. Fabian, A. C. Cooling flows in clusters of galaxies. *Annu. Rev. Astron. Astrophys.* **32**, 277-318 (1994).
5. Peterson, J. R. & Fabian, A. C. X-ray spectroscopy of cooling clusters. *Phys. Rep.* **427**, 1-39 (2006).
6. Churazov, E., Forman, W., Jones, C. & Böhringer, H. Asymmetric, arc minute scale structures around NGC 1275. *Astron. Astrophys.* **256**, 788-794 (2000).
7. McNamara, B. R. & Nulsen, P. E. J. Heating hot atmospheres with active galactic nuclei. *Annu. Rev. Astron. Astrophys.* **45**, 117-175 (2007).
8. Fabian, A. C. Observational evidence of active galactic nuclei feedback. *Annu. Rev. Astron. Astrophys.* **50**, 455-489 (2012).
9. Bîrzan, L. et al. The duty cycle of radio-mode feedback in complete samples of clusters. *Mon. Not. R. Astron. Soc.* **427**, 3468-3488 (2012).
10. Churazov, E., Brüggen, M., Kaiser, C. R., Böhringer, H. & Forman, W. Evolution of

buoyant bubbles in M87. *Astrophys. J.*, **554**, 261-273 (2001).

11. Omma, H., Binney, J., Bryan, G. & Slyz, A. Heating cooling flows with jets. *Mon. Not. R. Astron. Soc.* **348**, 1105-1119 (2004).

12. Hillel, S. & Soker, N. Heating cold clumps by jet-inflated bubbles in cooling flow clusters. Preprint at http://arxiv.org/abs/1403.5137.

13. Zhuravleva, I. et al. The relation between gas density and velocity power spectra in galaxy clusters: qualitative treatment and cosmological simulations. *Astrophys. J. Lett.* **788**, L13-L18 (2014).

14. Gaspari, M. et al. The relation between gas density and velocity power spectra in galaxy clusters: high-resolution hydrodynamic simulations and the role of conduction. *Astron. Astrophys.* **569**, A67–A82 (2014).

15. Fabian, A. C. et al. A very deep Chandra observation of the Perseus Cluster: shocks, ripples and conduction. *Mon. Not. R. Astron. Soc.* **366**, 417-428 (2006).

16. Sternberg, A. & Soker, N. Sound waves excitation by jet-inflated bubbles in clusters of galaxies. *Mon. Not. R. Astron. Soc.* **395**, 228-233 (2009).

17. Brethouwer, G., Billant, P., Lindborg, E. & Chomaz, J.-M. Scaling analysis and simulation of strongly stratified turbulent flows. *J. Fluid Mech.* **585**, 343-368 (2007).

18. Arévalo, P., Churazov, E., Zhuravleva, I., Hernández-Monteagudo, C. & Revnivtsev, M. A Mexican hat with holes: calculating low-resolution power spectra from data with gaps. *Mon. Not. R. Astron. Soc.* **426**, 1793-1807 (2012).

19. Dennis, T. J. & Chandran, B. D. G. Turbulent heating of galaxy-cluster plasmas. *Astrophys. J.* **622**, 205-216 (2005).

20. Kolmogorov, A. N. The local structure of turbulence in incompressible viscous fluid for very large Reynolds' numbers. *Dokl. Akad. Nauk SSSR.* **30**, 301-305 (1941).

21. Sreenivasan, K. R. On the universality of the Kolmogorov constant. *Phys. Fluids.* **7**, 2778-2784 (1995).

22. Kaneda, Y., Ishihara, T., Yokokawa, M., Itakura, K. & Uno, A. Energy dissipation rate and energy spectrum in high resolution direct numerical simulations of turbulence in a periodic box. *Phys. Fluids.* **15**, L21-L24 (2003).

23. Beresnyak, A. Spectral slope and Kolmogorov constant of MHD turbulence. *Phys. Rev. Lett.* **106**, 075001 (2011).

24. Schekochihin, A. A. et al. Astrophysical gyrokinetics: kinetic and fluid turbulent cascades in magnetized weakly collisional plasmas. *Astrophys. J. Suppl. S.* **182**, 310-377 (2009).

25. Churazov, E. et al. X-ray surface brightness and gas density fluctuations in the Coma cluster. *Mon. Not. R. Astron. Soc.* **421**, 1123-1135 (2012).

26. Sutherland, R. S. & Dopita, M. A. Cooling functions for low-density astrophysical plasmas. *Astrophys. J. Suppl. S.* **88**, 253-327 (1993).

27. Markevitch, M. & Vikhlinin, A. Shocks and cold fronts in galaxy clusters. *Phys. Rep.* **443**, 1-53 (2007).

28. Subramanian, K., Shukurov, A. & Haugen, N. E. L. Evolving turbulence and magnetic fields in galaxy clusters. *Mon. Not. R. Astron. Soc.* **366**, 1437-1454 (2006).

29. Balbus, S. A. & Soker, N. Resonant excitation of internal gravity waves in cluster cooling flows. *Astrophys. J.* **357**, 353-366 (1990).

**Acknowledgements** Support for this work was provided by the NASA through Chandra award number AR4-15013X issued by the Chandra X-ray Observatory Center, which is operated by the Smithsonian Astrophysical Observatory for and on behalf of the NASA under contract NAS8-03060. S.W.A. acknowledges support from the US Department of Energy under contract number DE-AC02-76SF00515. I.Z. and N.W. are partially supported from Suzaku grants NNX12AE05G and NNX13AI49G. P.A. acknowledges financial support from Fondecyt 1140304 and European Commission's Framework Programme 7, through the Marie Curie International Research Staff Exchange Scheme LACEGAL (PIRSES-GA -2010-2692 64). E.C. and R.S. are partially supported by grant no. 14-22-00271 from the Russian Scientific Foundation.

**Author Contributions** I.Z.: data analysis, interpretation, manuscript preparation; E.C.: data analysis, interpretation, manuscript preparation; A.A.S.: interpretation, discussions, manuscript preparation; A.F.: principal investigator of the Perseus Cluster observations, interpretation, manuscript review; S.A.: interpretation, discussions, manuscript review; W.F.: principal investigator of the M87 observations, interpretation, manuscript review; P.A., J.S., A.S., R.S., A.V., N.W.: interpretation, discussions and manuscript review.

**Author Information** Reprints and permissions information is available at www.nature.com/reprints. The authors declare no competing financial interests. Readers are welcome to comment on the online version of the paper. Correspondence and requests for materials should be addressed to I.Z. (zhur@stanford.edu).

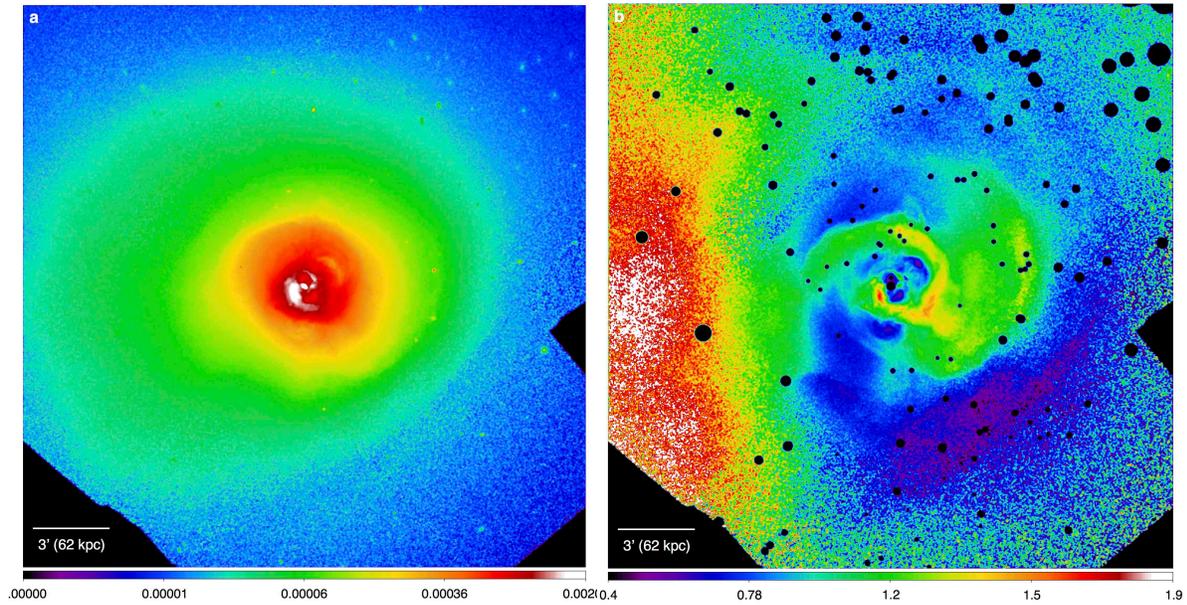

**Figure 1 | X-ray image of the core of the Perseus Cluster.** (a) X-ray surface brightness in units of counts/s/pixel obtained in the 0.5-3.5 keV energy band from Chandra observations. (b) The same divided by the mean surface-brightness profile, highlighting the relative density fluctuations. The images are smoothed with a 3″ Gaussian. Black circles: excised point sources (see Methods). The redshift is taken to be $z = 0.01756$ (redshift of the central galaxy); hence the angular diameter distance is 72 Mpc (for $h$=0.72, $\Omega_m$=0.3, $\Omega_\Lambda$=0.7) and 1' corresponds to a physical scale of 20.82 kpc.

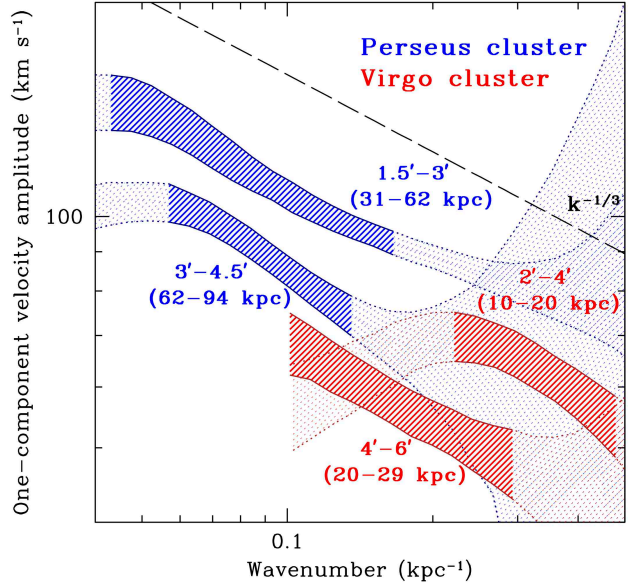

**Figure 2 | Measured amplitude of the one-component velocity $V_{1,k}$ of gas motions versus wavenumber $k$.** The amplitude is shown for two different annuli in both Perseus (blue) and M87/Virgo (red). The values are obtained from the power spectra of density fluctuations, derived from the X-ray images. The wavenumber $k$ is related to the spatial scale $l$ as $k=1/l$. Shaded regions show the range of scales where the measurements are robust against observational limitations (Methods). The width of each curve reflects the estimated $1\sigma$ statistical and stochastic uncertainties. The dashed line is the Kolmogorov scaling $k^{-1/3}$.

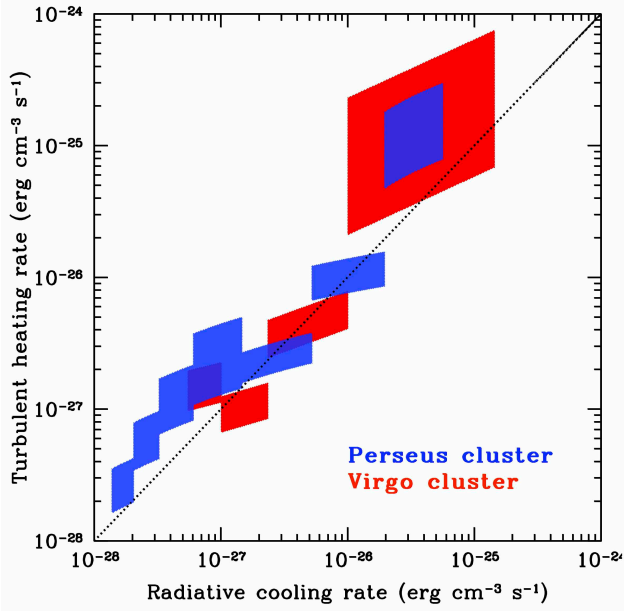

**Figure 3 | Turbulent heating ($Q_{heat}$) versus gas cooling ($Q_{cool}$) rates in the Perseus and Virgo cores.** Each shaded rectangle shows the heating and cooling rates estimated in a given annulus (top right – the innermost radius; bottom left – the outermost radius; see Extended Data Fig. 3). The size of each rectangle reflects 1$\sigma$ statistical and stochastic uncertainties in heating, variations of the mean gas density and temperature across each annulus (affecting estimates of both cooling and heating) and the deviations of the measured spectral slope from the Kolmogorov law.

## METHODS

**Data processing**

We use Chandra data ObsIDs: 3209, 4289, 4946 - 4953, 6139, 6145, 6146, 11713 - 11716, 12025, 12033 - 12037 for the Perseus Cluster and ObsIDs: 2707, 3717, 5826 - 5828, 6186, 7210 - 7212, 11783 for the Virgo Cluster to extract projected density fluctuation spectra in a set of radial annuli. The initial data processing has been done following the standard procedure[30], applying the most recent calibration data. To obtain the thermodynamic properties of both clusters, the spectra are deprojected[31] and fitted in the 0.6-9 keV band, using the XSPEC[32,33] code and APEC plasma model based on ATOMDB version 2.0.1. The spectral modeling approximates the emission from each shell as a single-temperature plasma in collisional equilibrium and assumes a constant metal abundance of 0.5 solar[34].

The X-ray mosaic image and its reduced counterpart for the Virgo Cluster are shown in Extended Data Fig. 2. The 0.5-3.5 keV band was used because it contains the dominant fraction of the cluster signal and because of the weak temperature dependence of the gas emissivity in this band. The image of relative fluctuations is obtained by dividing the mosaic image by a spherically symmetric $\beta$ model of the mean surface-brightness profile taking a core radius $0.34' = 1.7$ kpc and slope $\beta=0.39$. Point sources have been excised from the images, using circles scaled according to the size of the combined PSF. Extended Data Fig. 3 shows the set of annuli in Perseus and Virgo in which this analysis was performed.

**Mean profiles**

Deprojected radial profiles of the electron number density $n_e$ and temperature $T_e$ are shown in Extended Data Fig. 1 for both clusters. Note that the properties of the two clusters are very different. In particular, the density in Virgo is a factor of $\sim 3$ (or more) lower than in Perseus at radii beyond $\sim 10$ kpc. The temperature in Virgo is also lower, by a factor of $\sim 1.5 - 2$ at $r \sim 10 - 20$ kpc. Yet, $Q_{turb} \sim Q_{cool}$ in both clusters, as shown in Fig. 3, suggesting a self-regulated mechanism such as, e.g., the AGN feedback model[35].

The mean mass density of the gas is $\rho_0 = (n_e + n_i)\mu m_p = \xi \mu m_p n_e$, where $n_i=(\xi-1)n_e$ is the ion number density and $m_p$ is the proton mass. Consider a fully ionized plasma with an abundance of heavy elements $\sim 0.5$ Solar, $\xi=1.912$ and the mean particle weight $\mu=0.61$. The cooling time is defined as $t_{cool} = \frac{3}{2}\frac{(n_e+n_i)k_B T}{n_e n_i \Lambda_n(T)} = \frac{3}{2}\frac{\xi}{\xi-1}\frac{k_B T}{n_e \Lambda_n(T)}$, where $\Lambda_n(T)$ is the normalized cooling function[26] [erg cm$^3$ s$^{-1}$], $k_B$ is the Boltzmann constant, and we assume identical ion and electron temperatures: $T=T_e=T_i$. The sound speed, treating the ICM as an ideal monatomic gas, is $c_s = \sqrt{\frac{5}{3}\frac{k_B T}{\mu m_p}}$.

Both $t_{cool}$ and $c_s$ are plotted in Extended Data Fig. 1 as functions of radius. It is manifest that $t_{cool}$ is shorter than the Hubble time in the central $\sim 100$ kpc. Note that the cooling time is at least $\sim 7 - 20$ times longer than the characteristic free-fall time $t_{ff}$ in both clusters, defined in terms of the radius $r$ and the gravitational acceleration $g$ as $t_{ff}=(2r/g)^{1/2}$. Therefore, thermal instability is at most marginally important for the hot gas[36].

The cooling time $t_{cool}$ and the cooling rate $Q_{cool}$ have been calculated using a gas cooling function $\Lambda_n(T)$ with solar metallicity. This is a conservative choice, since the dependence of the cooling function on metallicity is not strong and often can be neglected for typical ICM gas temperatures $\sim 2 \cdot 10^7 - 10^8$ K. In addition, metallicity measurements in the cores of clusters from X-ray spectra can be biased due to the complexity of the spectral modeling of multi-temperature plasma. Accounting for radial metallicity variations in both clusters (based on the simplest one-temperature spectral model) and the consequent variation of the cooling function, the cooling rates shown in Fig. 3 and Extended Data Fig. 4 may be lower by a factor $\sim 0.8$ in Perseus and in the outermost annuli in Virgo, but higher by a factor of $\sim 2$ in the innermost annuli in Virgo.

**A priori estimates of velocity and density fluctuations required for heating-cooling balance**

It is useful to have *a priori* estimates of the fluctuation amplitudes required to make a heating-cooling balance plausible. Equating $Q_{cool} = n_e n_i \Lambda_n(T)$ and $Q_{turb} = C_Q \rho_0 V_{1,k}^3 k$, the characteristic Mach number of the turbulent motions at scale $l=1/k$ becomes:

$$\text{Ma} = \sqrt{3}\frac{V_{1,k}}{c_s} = \sqrt{3}\left(\frac{\xi-1}{\xi \mu m_p} \frac{\Lambda_n(T)}{C_Q}\right)^{1/3} n_e^{1/3} c_s^{-1} k^{-1/3} \approx$$

$$0.15 \left(\frac{n_e}{10^{-2} cm^{-3}}\right)^{1/3} \left(\frac{c_s}{1000 \ km \ s^{-1}}\right)^{-1} \left(\frac{l}{10 \ kpc}\right)^{1/3}.$$

Here we have referred all of the equilibrium quantities to their typical order-of-magnitude values and used the fact that the normalized cooling function $\Lambda_n(T)$ is a weak function of the ICM temperature[26], allowing us to adopt the mean value $\Lambda_n \approx 2.5 \cdot 10^{-23}$ erg cm$^3$ s$^{-1}$ (for a gas with solar metallicity). Since bubbles have typical sizes $\sim 5 - 20$ kpc[8], the value $l \sim 10$ kpc is a reasonable order-of-magnitude estimate of the outer scale for the ICM turbulence driven by such bubbles in cluster cores. Thus, the dissipation of turbulence with relatively low Mach numbers, Ma $\sim 0.15$, should be sufficient to balance the cooling of the gas in cores.

In view of the relationship $\delta \rho_k / \rho_0 \approx \eta_1 V_{1,k}/c_s$ between the amplitudes of density and velocity fluctuations[13], these Mach numbers correspond to $\delta \rho / \rho_0 \sim 10\%$. These are indeed typical values of density fluctuations we see in galaxy clusters.

**Trivial part of the correlation between heating and cooling**

As the density explicitly enters the expressions for both the cooling rate and turbulent heating rate, the linear correlation between these rates seen in Fig. 3 partly reflects the large range of mean densities at different radii (Extended Data Fig. 1). In order to show that the correlation is not due *solely* to this trivial part, we divide both $Q_{cool}$ and $Q_{turb}$ by the density $\rho_0$ and thus obtain the cooling and heating rates per unit mass [erg s$^{-1}$g$^{-1}$], see Extended Data Fig. 4. Although the range of values of both rates is now smaller, as expected, the correlation between them remains manifest.

**Systematic uncertainties in the measurement of density-fluctuation amplitudes**

We start with the measurements of the surface-brightness fluctuations based on broad-band X-ray images[25] (details in I.Z. *et al.*, manuscript in preparation), using the $\Delta$-variance method[18,37]. The variance at scale $l$ estimated using this method corresponds to a convolution of the original power spectrum with a broad filter. For a Kolmogorov-like power spectrum, the method can overestimate[18] the amplitude of fluctuations by ~25%.

A more important source of uncertainties in the determination of the density power spectrum is the fact that dividing the cluster image into "perturbed" and "unperturbed" components is ambiguous, especially for a relatively steep perturbation spectrum like Kolmogorov's, whose integrated power is dominated by the largest scales[25,38]. The $\beta$ model provides a reasonable description of the radial surface-brightness profiles for Perseus and Virgo. It is, therefore, a sensible starting choice of an unperturbed cluster model. Of course, more complicated models, e.g., projection of an ellipsoidal $\beta$ model or models with more sophisticated radial profiles, could be used as well. Adding more flexibility (more fitting parameters) to the model allows one to absorb more large-scale features of the image into the model surface-brightness distribution. The net result of such improved fitting is that the measured power in the remaining perturbations will decrease on large scales, while the small-scale power will be less affected (provided the spectrum, $E(k)$, is not steeper than $k^{-3}$, which would correspond to the spectral tail of a smooth large-scale distribution; indeed, all our measured spectra are close to the Kolmogorov $k^{-5/3}$ spectrum, which satisfies this constraint). This would cause the power spectrum to flatten at large scales. This model-dependent nature of the large scales is a feature of any division of the surface-brightness variations into "unperturbed" and "perturbed" parts, including the case of the simplest $\beta$ model. This is why we expect that the estimates of the heating power based on small-scale tail in the inertial range are likely more robust than estimates based on larger, outer scales. Our estimate of $\varepsilon$ is, thus, not very susceptible to the choice of the underlying model of the mean surface-brightness profile.

The reconstruction of the three-dimensional power spectrum of density fluctuations $P_{3D}$ from the two-dimensional power spectrum of the surface-brightness fluctuations $P_{2D}$ is another source of uncertainty. The geometrical factor $f_{2D \rightarrow 3D}=P_{2D}/P_{3D}$ depends on the radial profile of the surface brightness[25]. We use the mean value of $f_{2D \rightarrow 3D}$ for each annulus and conservatively estimate the uncertainties by comparing it with the factors for the inner and outer radii of the same annulus. The maximal uncertainty does not exceed 20% except for the innermost region of M87/Virgo.

The random nature of density fluctuations is another source of uncertainty. The spectra we calculate are based on squared amplitudes averaged over each annulus. Given a (expected) large degree of intermittency of density fluctuations and a limited spatial extent of the annuli, one might ask how representative and how statistically converged such annular averages are. For example, analyzing fluctuations in small patches within the 3′ - 4.5′ (62 - 94 kpc) annulus in Perseus, we find $\delta\rho_k/\rho_0$ at scales $k^{-1} \approx$ 15 kpc varying in a relatively broad range from 3% to 10%. This difficulty in relating the rms turbulence level to what happens (and what is observed) in any given location is unavoidable as one always observes only a single realization of the fluctuating field. In order to achieve statistical convergence, we perform our averages in relatively wide annuli. The results we report are robust in the sense that choosing twice broader annuli does not change the conclusions.

A related problem is associated with the weighting scheme used to calculate the amplitude of the fluctuations within each annulus by averaging an image after applying a filter that selects perturbations with a given spatial scale. The exposure maps of the images are not uniform and the brightness of the cluster itself also varies substantially across each annulus. The optimal weighting scheme for the reduction of Poisson noise would require the weights to be $w_1 \sim t_{exp}I_0$, where $t_{exp}$ is the exposure map and $I_0$ is the global $\beta$-model profile of the surface brightness. This means that those parts of the cluster that have higher numbers of counts would have larger weights. We have experimented with two other choices of weights: $w_2 \sim t_{exp}$ and $w_3=1$. These weights have larger statistical errors, but provide a more uniform scheme for evaluating the amplitudes of the surface brightness fluctuations across the image. For the analysis reported in this Letter, we used the uniform weight $w_3$=1. In most cases (except for the innermost regions of the two clusters), the uncertainty associated with the choice of the weights does not exceed 20%.

The vertical width ("error bars") of the spectra shown in Fig. 2 and Extended Data Fig. 4 reflects the 1σ statistical uncertainty. The uncertainties discussed above slightly affect the shape of the spectra and may change the normalization by the factors estimated above (for details see I.Z. *et al.*, manuscript in preparation). The dark-shaded regions of the spectra in Fig. 2 and Extended Data Fig. 4b show the wavenumber ranges over which we deem the spectra to be determined reliably – these ranges were used to determine the turbulent cascade rate $\varepsilon$ in the manner described in the main text. The high-$k$ limits of these ranges are set by the "statistical" uncertainty (Poisson noise) and/or by the PSF distortions of the amplitude (in both cases the uncertainty is less than 20% in the "reliable" range). At low $k$, we limit our "reliable" $k$ ranges by the wave numbers where the spectra start flattening. The shape of the spectra at these scales is most likely determined by the presence of several characteristic length scales (e.g., distance from the cluster center, scale heights) and by the large-scale uncertainties inherent in the choice of the underlying model of the "unperturbed" cluster and in using finite-width annular averaging regions. This flattening disappears or shifts to smaller $k$ if thicker annuli are used.

**Systematic uncertainties in the conversion of density-fluctuation amplitudes to velocity amplitudes**

If the perturbations of the intracluster gas are small, one expects a linear relationship between the velocity $V_{1,k}$ and density $\delta\rho_k/\rho_0$ spectral amplitudes[13], $\frac{\delta\rho_k}{\rho_0} = \eta_1 \frac{V_{1,k}}{c_s}$, with $\eta_1 \sim 1$ set by gravity-wave physics. This assumes that the injection scale of the turbulence is larger than or comparable to the Ozmidov scale[39] – the scale at which the turbulent eddy turnover time scale becomes smaller than the buoyancy (Brunt–Väisälä) time scale (i.e., nonlinear advection becomes more important than the buoyancy response). Dimensionally, this scale is $l_O = N^{-3/2}\varepsilon^{1/2}$, where $N = c_s/H$ is the Brunt-Väisälä frequency ($H$ is the hydrostatic equilibrium scale height – we have omitted numerical factors and ignored the distinction between entropy, pressure and temperature scale heights) and $\varepsilon = Q_{turb}/\rho_0$ is the turbulent cascade rate. The relationship $\eta_1 \sim 1$ is inherited from large scales at all scales $l<l_O$, where the density becomes a passive scalar[13].

Assuming that radiative cooling is balanced by turbulent heating, $Q_{turb}=Q_{cool}$, it is possible to make an *a priori* estimate of $l_O$ by letting $\varepsilon = Q_{cool}/\rho_0$ and using the local mean thermodynamic properties of the ICM to calculate $Q_{cool}$, $\rho_0$ and $N$. We have done this for both clusters, for each of the annuli where we subsequently calculated $Q_{turb}$ (Extended Data Fig. 4). In all cases, $l_O$ is within the range of scales (in some cases, comparable to the largest scales) over which velocity amplitudes were measured and used to calculate $Q_{turb}$, and for which the conclusion that $Q_{turb} \sim Q_{cool}$ was drawn. Therefore, our assumption of $\eta_1 = 1$ is at least self-consistent.

This assumption is also restricted to the inertial range, i.e., to scales larger than any dissipative cutoffs. It is interesting to compare the smallest scales that we are probing with the Kolmogorov (dissipative) scale $l_K = v^{3/4}/\varepsilon^{1/4}$, where $v$ is the kinematic viscosity calculated for unmagnetized gas (which is approximately the same as the parallel viscosity for a magnetized plasma[40]). In all regions considered in this work, the Kolmogorov scale is significantly smaller than the smallest scale used by us for the determination of the cascade rate. In the regions shown in Extended Data Fig. 4, $l_K \sim 0.5$ and 2 kpc ($k_K \sim 2$ and 0.5 kpc$^{-1}$) in the 1.5′ - 3′ and 3′ - 4.5′ annuli in Perseus, respectively. In the Virgo Cluster, $l_K \sim 0.3$ and 0.8 kpc ($k_K \sim 3$ and 1.3 kpc$^{-1}$) in the 2′ - 4′ and 4′ - 6′ respectively.

Cosmological simulations of galaxy clusters confirm that $\eta_1 \approx 1$ with a scatter of 30%[13]. Hydrodynamic simulations with controlled driving of turbulence also show $\eta_1 \approx 1$, provided thermal conduction is suppressed[14]. The 30% scatter in the value of $\eta_1$ gives a factor of 0.3 - 2 uncertainty in the heating rate.

We conclude that the cumulative uncertainty in the estimated heating rate is about a factor of ~ 3. While this uncertainty is large, the approximate agreement between heating and cooling rates is an interesting result, reinforced by the fact that not only numerically the two rates are comparable but also linearly correlated with each other. A more rigorous test will become possible with direct measurements of the velocity field by future X-ray observatories.

**Theoretical uncertainties: comments on the nature of ripples in the Perseus Cluster and on ICM heating theories**

Un-sharp-masking of the Perseus image shows rough concentric rings, so-called "ripples", in the surface brightness[15]. The observed morphology of these features, namely narrow in the radial direction and wide in the azimuthal direction, suggests two plausible possibilities: concentric sound waves[15] or stratified turbulence[13,17]. In the first case, the radial scale of the ripples should be determined by the time variability of the central AGN activity (intervals between outbursts, multiple-sound-wave excitation by vortices arising during each bubble inflation episode[16], also distance from the center, ICM properties etc.). In contrast, in the case of stratified turbulence, the radial scale $\Delta r$ will be determined by the ratio of the characteristic scale height $H$ in the atmosphere and the velocity amplitude $V$, viz., $\Delta r \sim HV/c_s$. Here we assume the second scenario and defer the detailed analysis of the nature of the substructure to a future publication.

Many other models of ICM heating, that could in principle offset radiative cooling in cluster cores, have been suggested. They differ widely in (i) their presumed primary source of energy and (ii) in how this energy is channeled to the ICM. A brief and incomplete list of the broad classes into which these models fall is as follows:

1) source: thermal energy of the cluster gas; channeling mechanism: conductive heat flux to the core[42,43],

2) source: cluster mergers; channeling mechanism: turbulence[28,44],

3) source: galaxy motions; channeling mechanism: turbulence[28,29,45,46],

4) source: central AGN; channeling mechanism: shocks and sound waves[15,47], turbulent dissipation[48,49], turbulent mixing[50], cosmic rays[51,52], radiative heating[53,54], mixing of gas between ICM and the hot content of bubbles[12], etc.

Given the multiplicity of possible scenarios, a detailed discussion and comparison of these models or even a complete list of references are beyond the scope of this Letter. We refer the reader to review papers 7,8 and references therein. The content of this Letter is focused on the energy channeling mechanism rather than the energy source. Note that along with turbulent dissipation, turbulent heat conduction might also play a role in the cooling-heating balance. It can be shown, however, that in cluster cores and assuming either stratified or isotropic turbulence, its contribution cannot be much larger than that of the turbulent heating (A.A.S. *et al.*, manuscript in preparation).

30. Vikhlinin, A. et al. Chandra temperature profiles for a sample of nearby relaxed galaxy clusters. *Astrophys. J.* **628**, 655-672 (2005).

31. Churazov, E., Forman, W., Jones, C. & Böhringer, H. XMM-Newton observations of the Perseus Cluster. I. The temperature and surface brightness structure. *Astrophys. J.* **590**, 225-237 (2003).

32. Foster, A. R., Ji, L., Smith, R. K. & Brickhouse, N. S. Updated atomic data and calculations for X-ray spectroscopy. *Astrophys. J.* **756**, 128-139 (2012).

33. Smith, R. K., Brickhouse, N. S., Liedahl, D. A. & Raymond, J. C. Collisional plasma models with APEC/APED: emission-line diagnostics of hydrogen-like and helium-like Ions. *Astrophys. J. Lett.*, **556**, L91-L95 (2001).

34. Anders, E. & Grevesse, N. Abundances of the elements - meteoritic and solar. *Geochim. Cosmochim. Ac.* **53**, 197-214 (1989).

35. Churazov, E., Sunyaev, R., Forman, W. & Böhringer, H. Cooling flows as a calorimeter of active galactic nucleus mechanical power. *Mon. Not. R. Astron. Soc.* **332**, 729-734 (2002).

36. McCourt, M., Sharma, P., Quataert, E. & Parrish, I. J. Thermal instability in gravitationally stratified plasmas: implications for multiphase structure in clusters and galaxy haloes. *Mon. Not. R. Astron. Soc.* **419**, 3319-3337 (2012).

37. Ossenkopf, V., Krips, M. & Stutzki, J. Structure analysis of interstellar clouds. I. Improving the Δ-variance method. *Astron. Astrophys.* **485**, 917-929 (2008).

38. Sanders, J. S. & Fabian, A. C. Deep Chandra and XMM-Newton X-ray observations of AWM 7 - I. Investigating X-ray surface brightness fluctuations. *Mon. Not. R. Astron. Soc.* **421**, 726-742 (2012).

39. Ozmidov, R. V. Length scales and dimensionless numbers in a stratified ocean. *Oceanology* **32**, 259–262 (1992).

40. Braginskii, S. I. Transport processes in a plasma. *Rev. Plasma Phys.* **1**, 205 (1965).

41. Werner, N. et al. XMM-Newton high-resolution spectroscopy reveals the chemical evolution of M87. *Astron. Astrophys.* **459**, 353-360 (2006).

42. Zakamska, N. L. & Narayan, R. Models of galaxy clusters with thermal conduction. *Astrophys. J.* **582**, 162-169 (2003).

43. Cho, J. et al. Thermal conduction in magnetized turbulent gas. *Astrophys. J. Lett.* **589**, L77-L80 (2003).

44. Norman, M. L. & Bryan, G. L. Cluster turbulence. *Proc. The Radio Galaxy Messier 87* **530**, 106-115 (1999).

45. Lufkin, E. A., Balbus, S. A. & Hawley, J. F. Nonlinear evolution of internal gravity waves in cluster cooling flows. *Astrophys. J.* **446**, 529-540 (1995).

46. Ruszkowski, M. & Oh, S. P. Galaxy motions, turbulence and conduction in clusters of galaxies. *Mon. Not. R. Astron. Soc.* **414**, 1493-1507 (2011).

47. Randall, S. W. et al. Shocks and cavities from multiple outbursts in the galaxy group NGC 5813: a window to active galactic nucleus feedback. *Astrophys. J.* **726**, 86-104 (2011).

48. Fujita, Y., Matsumoto, T. & Wada, K. Strong turbulence in the cool cores of galaxy clusters: can tsunamis solve the cooling flow problem? *Astrophys. J.* **612**, L9-L12 (2004).

49. Banerjee, N. & Sharma, P. Turbulence and cooling in galaxy cluster cores. *Mon. Not. R. Astron. Soc.* **443**, 687-697 (2014).

50. Kim, W.-T. & Narayan, R. Turbulent mixing in clusters of galaxies. *Astrophys. J.* **596**, L139-L142 (2003).

51. Chandran, B. D. & Dennis T. J. Convective stability of galaxy-cluster plasmas. *Astrophys. J.* **642**, 140-151 (2006).

52. Pfrommer, C. Toward a comprehensive model for feedback by active galactic nuclei: new insights from M87 observations by LOFAR, Fermi, and H.E.S.S. *Astrophys. J.* **779**, 10-28 (2013).

53. Ciotti, L. & Ostriker, J. P. Cooling flows and quasars. II. Detailed models of feedback-modulated accretion flows. *Astrophys. J.,* **551**, 131-152 (2001).

54. Nulsen, P. E. J. & Fabian, A. C. Fuelling quasars with hot gas. *Mon. Not. R. Astron. Soc.* **311**, 346-356 (2000).

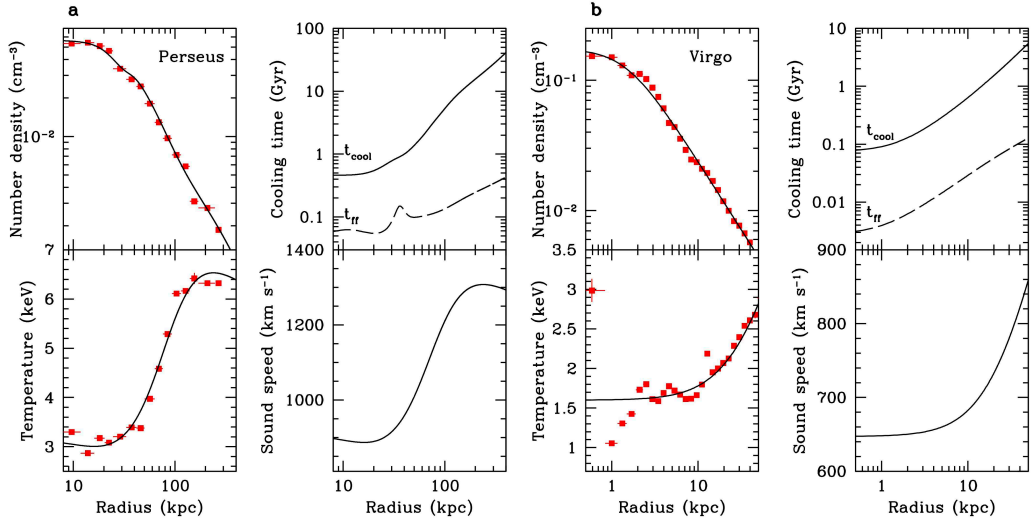

**Extended Data Figure 1 | Thermodynamic properties of the Perseus and Virgo Clusters.** Radial profiles of the deprojected electron number density, the electron temperature, the cooling ($t_{cool}$) and the free-fall ($t_{ff}$) times, and the sound speed. Red points: data with 1σ error bars (s.d.); black curves: data approximations by smooth functions. The increased temperature scatter in the central few kpc is associated with the presence of multi-temperature plasma in cool cores. A two-temperature fit of high-resolution XMM-Newton RGS spectra of the core of Virgo suggests an ambient temperature there of ~1.6 keV[41]. The smooth functional approximation we have chosen therefore approaches this value.

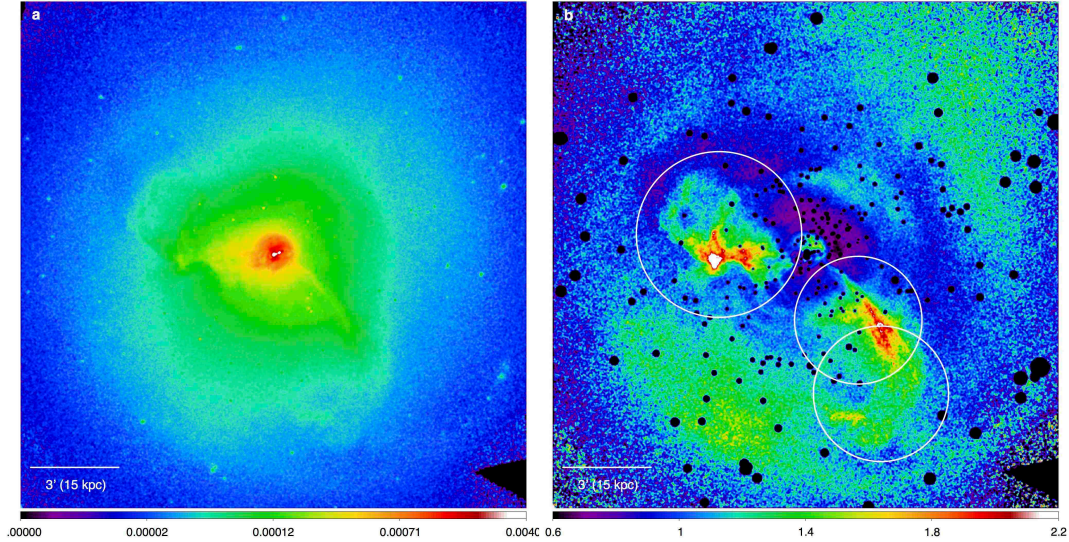

**Extended Data Figure 2 | X-ray image of the core of the Virgo Cluster.** (a) X-ray surface brightness in units of counts/s/pixel in the 0.5-3.5 keV energy band. (b) Relative surface brightness fluctuations. Both images are smoothed with a 3″ Gaussian. Black circles: excised point sources and central jet. White circles indicate "arm-like" structures associated with the central AGN's activity, which have also been excised. We adopt 16.9 Mpc as the distance to the cluster, implying than an angular size of 1′ corresponds to a physical scale of 4.91 kpc.

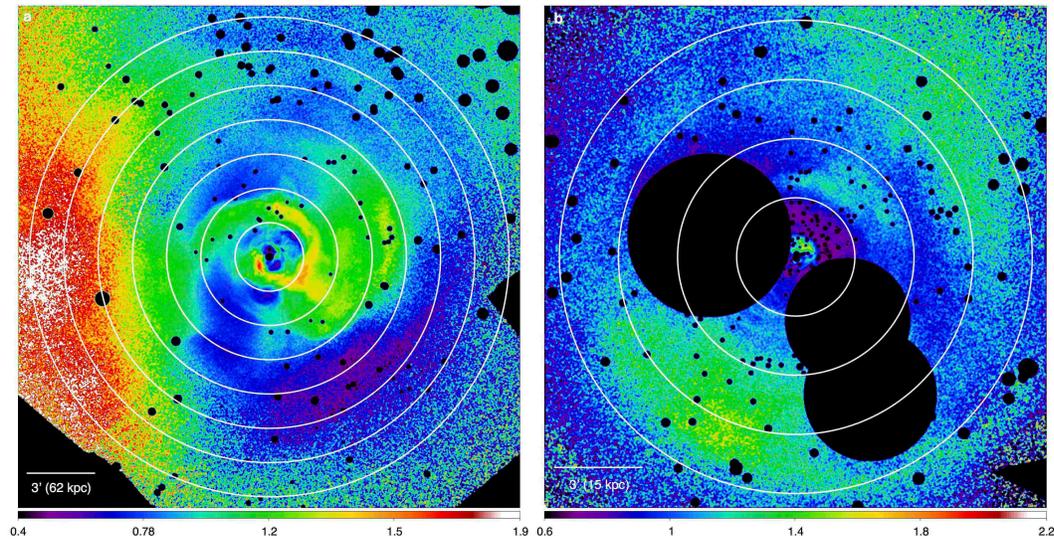

**Extended Data Figure 3 | Set of the radial annuli used in the analysis of the Perseus and Virgo clusters.** The same as panels (b) in Fig. 1 and Extended Data Fig. 1 with white circles indicating the annuli used. The width of each annulus is $1.5′ \approx 31$ kpc in Perseus (a) and $2′ \approx 9.8$ kpc in Perseus and Virgo (b). The outermost circles are $10.5′ \approx 218$ kpc and $8′ \approx 39$ kpc in Perseus and Virgo, respectively.

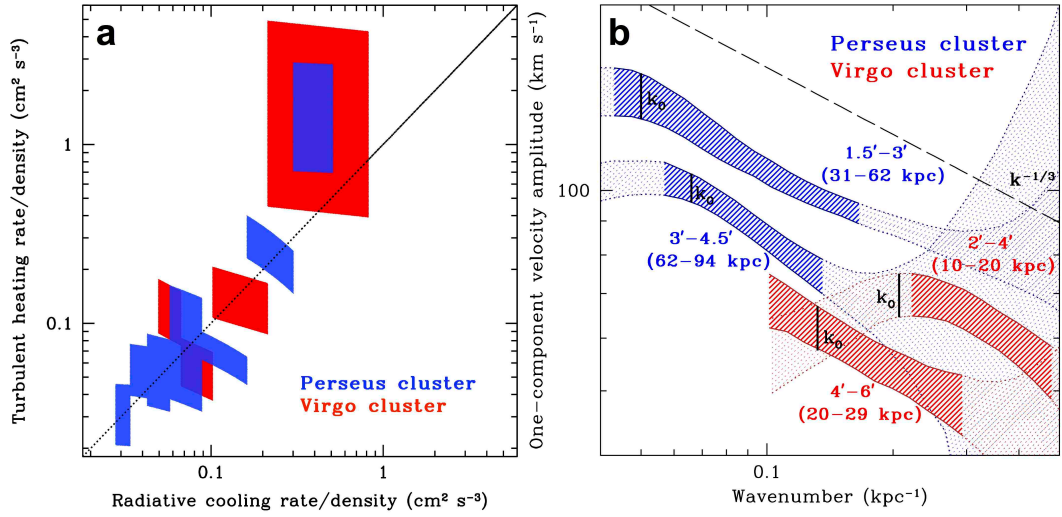

**Extended Data Figure 4 | Turbulent heating per unit density versus radiative cooling per unit density and the Ozmidov scale in the Perseus and Virgo clusters.** (a) The same as Fig. 3, but with the turbulent heating and cooling rates divided by the mass density of gas in each annulus. (b) The same as Fig. 2 with the Ozmidov scale $l_O = 1/k_O = N^{-3/2}\varepsilon^{1/2}$ shown for each annulus (vertical black lines), estimated with $\varepsilon = Q_{cool}/\rho_0$ (assuming $Q_{turb} = Q_{cool}$).